\documentclass[%
10pt,
aps,
prl,
nofootinbib,
nobibnotes,
amsmath,
amssymb,
twocolumn
]{revtex4-2}
\usepackage{amssymb}
\usepackage{physics}
\usepackage{graphicx}
\usepackage{import}
\usepackage{placeins}
\usepackage{siunitx}
\usepackage{amsmath}
\usepackage{amsthm}
\usepackage{amsfonts}
\usepackage{bbm}
\usepackage[normalem]{ulem}
\usepackage{soul}
\graphicspath{{./imgs}}

\usepackage{xcolor}
\usepackage[colorlinks]{hyperref}
\usepackage[caption=false, subrefformat=parens]{subfig}

\newcommand{\oldtext}[1]{\textcolor{purple}{\sout{#1}}}

\usepackage{bbold}
\usepackage{comment}
\renewcommand{\oldtext}[1]{} %

\newcommand\rA{r_\mathrm{A}}
\newcommand\rN{r_\mathrm{N}}
\renewcommand\L{-}

\newcommand\R{+}
\newcommand\EB{E_\mathrm{B}}

\newcommand\Veff{V_\mathrm{eff}}

\newcommand\chiral[1]{{#1_{\mathrm{c}}}}

\renewcommand\section[1]{\emph{#1}.---}

\begin{document}
\title{Quantum Hall transformer in a quantum point contact over the full range of transmission}

\author{Stuart Yi-Thomas}
\email[]{snthomas@umd.edu}
\affiliation{Condensed Matter Theory Center and Joint Quantum Institute, Department of Physics, University of Maryland, College Park, Maryland 20742-4111, USA}
\author{Jay D. Sau}
\affiliation{Condensed Matter Theory Center and Joint Quantum Institute, Department of Physics, University of Maryland, College Park, Maryland 20742-4111, USA}

\date{October 27, 2026}

\begin{abstract}
  A recent experiment [Cohen {\it et al.}, Science {\bf 382}, 542 (2023)] observed a robustly quantized $e^2/2h$ conductance in a quantum point contact between fractional quantum Hall edges (a quantum Hall transformer), which then vanishes at low temperature.
  While this behavior can be described by a boundary sine-Gordon (BSG) model derived from electron tunneling, the microscopic motivation for such strong tunneling is unclear. 
  We use an alternative model based on sliding charge density waves and a density inhomogeneity to clarify the BSG description and calculate the conductance over the full range of transmission.
  Using a perturbative method and a matrix product state calculation, we draw a quantitative connection between the BSG model and the physical quantum Hall system.
\end{abstract}


\maketitle

Fractional quantum Hall (FQH) systems are one of the few experimentally known platforms that support chiral Luttinger liquid edge states~\cite{chang2003chiral}.
Interestingly, a recent experiment~\cite{cohen2023universal} in graphene FQH quantum point contacts (QPCs) has seen differential conductance that closely matches theoretical predictions~\cite{chamon1997distinct,sandler1999noise,sandler1998andreev,kane1998resonant,chklovskii1998consequences} (power-law scaling at both low voltage and temperature as well 
as high-voltage universality) based on the boundary sine-Gordon (BSG) model~\cite{fendley1994exact}.
Such quantitative experiments, if truly described by the BSG model, open more vistas into its physics which can include rather intricate phenomena~\cite{fendley1994exact} accessed in relatively few experiments~\cite{anthore2018circuit}.
Additionally, the high-voltage universal regime of the BSG model, where the differential conductance is near a quantized value of $e^2/2h$~\cite{cohen2023universal}, realizes a dissipationless dc voltage transformer known as the quantum Hall transformer (QHT)~\cite{chklovskii1998consequences}.

\begin{figure}[t]
\centering
\includegraphics[width=1.0\columnwidth]{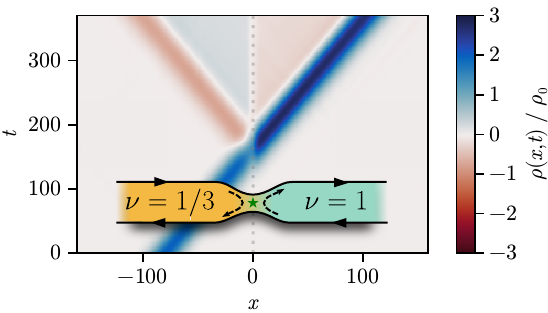}
\caption{\label{fig:trajectory} 
  Sign-flipping reflection of a charge wavepacket from the interface between $\nu=1/3$ and $\nu=1$ quantum Hall edge states and a backscattering impurity (green star),
  illustrating the quantum Hall transformer behavior. 
The local charge $\rho(x,t)$ at position $x$ and time $t$ is normalized by an arbitrary charge $\rho_0$ chosen to illustrate the $2e^*\rightarrow -e^*$ charge transfer where $3e^*=e$.
}
\end{figure}

Despite experimental agreement with the BSG model, the microscopic connection with the experiment remains unclear.
The original interpretation of the BSG model in the FQH context involves a tunneling term between the chiral Luttinger liquid states of $\nu=1/3$ and $\nu=1$ FQH edges~\cite{chamon1997distinct}, however such a perturbation is irrelevant under the renormalization group (RG)~\cite{kane1992transmission}, generically flowing to zero conductance. 
Additionally, the high-temperature scaling exponent of the conductance disagrees with the BSG prediction, agreeing instead with the fixed point in Ref.~\onlinecite{chklovskii1998consequences}.
Though several arguments~\cite{kane1998resonant,sandler1998andreev,sandler1999noise} have been proposed for how strong coupling behavior might arise in the BSG description,  none describe the junction over the full range of transmission or match the observed high-temperature exponent.

An alternative configuration for generically creating strong coupling~\cite{chklovskii1998consequences} (i.e.~the QHT limit) is in principle based on pinching the QPC to a width comparable to the magnetic length and then smoothly varying the density between the two FQH states at filling $\nu=1/3$ and $\nu=1$.
In such a QPC, the low density electron gas forms a sliding charge density wave (CDW) due to Coulomb interactions~\cite{fukuyama1978dynamics}.
This density wave is described by a conventional Luttinger liquid that is adiabatically connected to the pair of chiral edge modes far from the QPC.
In contrast to the tunneling term model \cite{chamon1997distinct} which uses two chiral Luttinger liquids, the density wave model uses a single \emph{non-chiral} Luttinger liquid.
 
The inhomogeneous electrostatic potential of the QPC then pins this CDW and is a relevant perturbation that prevents charge transport in the low-temperature limit~\cite{safi1999b,sedlmayr2012transport}.
Therefore this model provides a microscopic framework to describe both the high-temperature/voltage sliding limit with a differential conductance of $e^2/2h$~\cite{safi1995,chklovskii1998consequences} and the low-temperature/voltage pinned limit with vanishing differential conductance.
The pinning description of the QPC provides a controlled model for the crossover from the ballistic QHT to the insulating regime via the RG flow of the pinning potential.
However, the transport properties of this sliding CDW in the QPC and its connection to the BSG have not been studied so far.

In this Letter, we study this sliding CDW using the non-chiral Luttinger liquid with a pinning potential.
  First, we perturbatively calculate the conductance in the weak pinning limit.
  The result provides an experimentally-accessible measurement of the energy scale corresponding to the finite width of the junction in terms of the temperature-dependent dc conductance in the high-temperature limit.
  Next, we explore the ac conductivity at zero temperature using a matrix product state simulation, demonstrating both the QHT and zero-conductance limits.
  Then, we show that the Luttinger liquid model is described by the BSG model in the low temperature and frequency regime. 
  We fit the exact solution of this model to the matrix product state results, drawing a quantitative connection between the BSG parameters 
  and the pinning potential.
Finally, we discuss the case of a wide QPC and argue that a percolating channel emulates a narrow junction, justifying the Luttinger liquid model as an explanation for the results in Ref.~\onlinecite{cohen2023universal}.

\newcommand\CHg{g}
An impurity in a Luttinger liquid~\cite{kane1992transmission, safi1999b, fukuyama1978dynamics},  assuming the 
pinning potential~\cite{sedlmayr2012transport} to be of strength $\CHg$ at $x=0$, 
can be represented by the Hamiltonian %
\begin{equation}\label{eq:ham}
    H=\int \mathrm dx \frac{v(x)}{2\pi}\left[ K(x)\Pi^2+K(x)^{-1}(\partial_x\phi)^2\right] - \CHg \cos{2\phi(0)},
\end{equation}
where the boson field $\phi$ is defined in terms of the coarse-grained (on the scale of the inter-electron spacing) charge density $\rho=-\pi^{-1}\partial_x\phi$ in units of the electron charge $e$. 
Here $\Pi(x)$ is the momentum that is canonically conjugate (i.e.~$[\phi(x),\Pi(x')]=i\pi\delta (x-x')$)
to the boson field $\phi(x)$. %
The pair of edges of a FQH strip at filling fraction $\nu$ has a Luttinger parameter $K(x)=\nu$ and a mode velocity $v(x)$ that matches the edge velocity, which we take to be $1$ across the junction by rescaling $x$~\cite{sandler1998andreev}. 
To represent the junction, we smoothly vary $K(x)$ ~\cite{safi1999b}.
Conservation of charge requires that the current operator is written as $j=\pi^{-1}\partial_t\phi=(i/\pi)[H,\phi]=vK\Pi/\pi$.
The Hamiltonian in Eq.~\ref{eq:ham} is comparable to that of Ref.~\onlinecite{chklovskii1998consequences} in which $\CHg$ is a backscattering in the QPC, though the physics at finite $\CHg$ has yet to be explored in either model.

\section{Weak pinning}
\begin{figure}[t]
\centering
\includegraphics[width=1.0\columnwidth]{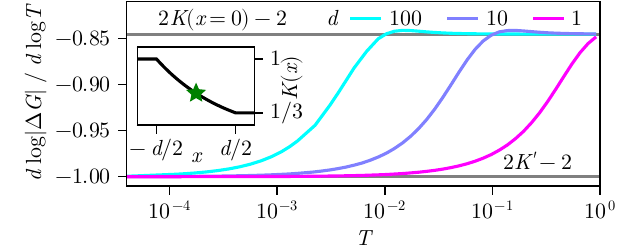}
\caption{\label{fig:perturbative_temperature_conductance} Scaling exponent of the perturbative correction to harmonic conductance for junctions widths $d$, calculated as the log-log slope of $|\Delta G(T)|$. The horizontal lines denote the low- and high-temperature limits, given by the effective Luttinger parameter $K'$ and the Luttinger parameter at the impurity $K(x=0)$. \textbf{(inset)} The spatial profiles of the Luttinger parameter $K(x)$---which has a piecewise exponential form--- with the backscattering impurity denoted by a green star.
}
\end{figure}

In the absence of pinning (i.e.\ $\CHg=0$), i.e.\ the quantum Hall transformer limit~\cite{chklovskii1998consequences},
the Hamiltonian in Eq.~\ref{eq:ham} is harmonic and the conductance $G_0$ is independent of temperature. 
By approximating the $K(x)$ profile as exponential over a region of width $d$ and constant otherwise (see Fig.~\ref{fig:perturbative_temperature_conductance} inset), the ac conductivity is analytically solvable by matching the boundary conditions of the piecewise solutions, yielding 
\begin{equation}
\label{eq:full-harmonic-conductance}
  G_0(\omega) = \frac{e^2}{h} \frac{\sqrt{K_\L K_\R}}{\cos \tilde\omega d - (i \omega/\tilde\omega) \sin \tilde\omega d},
\end{equation}
where $\tilde \omega^2 = \omega^2 - \left[ \frac 1 {2d} \log(K_\R/K_\L)\right]^2$ and $K_\pm = K(\pm \infty)$ (in this work, either 1 or 1/3).
See the Supplemental Material (SM)~\cite{supplemental} for the full derivation.
Note that since the velocity and $\hbar$ are set to one, the length $d$ has units of inverse energy.
This formula reduces to the quantum Hall transformer dc conductivity $G= K' e^2/h$ where $K'=2/(K_\L^{-1} + K_\R^{-1})=1/2$ is the effective Luttinger parameter~\cite{chamon1997distinct,chklovskii1998consequences} as $\omega$ goes to $0$.%

The $e^2/2h$ dc conductance in this limit can be understood as an Andreev--reflection-like process (see Fig.~\ref{fig:trajectory}) where a charge packet of charge $2e/3$ is reflected to a packet of $-e/3$ with the opposite sign. 
This process is a result of the conservation of chiral charge in the Luttinger liquid~\cite{wang2024interaction} 
in the absence of pinning 
since the chiral charge of the incoming and reflected packets are $K_-^{-1}\times 2e/3=2e$ and $K_-^{-1}\times e/3=e$ respectively while the transmitted packet has a chiral charge of $K_+^{-1} e=e$.

In the low-density sliding charge density wave picture, the pinning term $\CHg \cos 2\phi(0)$ can be understood as an oscillatory contribution to the charge density on top of the smooth background $\rho(x)$,
reflecting the spacing of the electrons which have a periodicity $\rho(x)^{-1}$
~\cite{fukuyama1978dynamics}.
This backscattering perturbation is known to be relevant in the RG sense~\cite{giamarchi2003quantum,chklovskii1998consequences} and suppresses the dc conductance, even from perturbatively small strengths $\CHg$, from the ideal transformer value (i.e.\ Eq.~\ref{eq:full-harmonic-conductance} at $\omega=0$) to zero at vanishing temperatures.
On the other hand, at high temperatures, the conductance suppression $\Delta G$ can be perturbatively analyzed in terms of the weak backscattering strength $\CHg$. Such an expansion provides a direct approach to define and measure $\CHg$ experimentally. 

We can determine the effect of a weak impurity on $\Delta G$ by performing a second-order perturbation theory calculation using the piecewise-exponential form of $K(x)$.
The conductance $G$, in the presence of pinning is computed as the linear response of the current $j$ to a voltage perturbation $\pi^{-1} V(t) \phi(0)$. 
We calculate the correction $\Delta G(T)$ to the dc conductance by taking the zero-frequency limit of the current due to an external voltage. The resulting conductance correction, which is calculated from the equations of motion for $\phi(x,t)$ under the Hamiltonian $H$ in Eq.~\ref{eq:ham} (see SM~\cite{supplemental} for derivation), is written as
  $\Delta G(T) = - (4 e^2/h)  g^2 \pi {K'}^2 \int_0^\infty dt \, t \; \mathcal G(t ) e^{-2\mathcal G_\text{S}(t,T)}$
where $\mathcal G(t)\equiv \mathcal G(0,0;t)$ is the retarded Green's function at $x=0$ and
$\mathcal G_\text{S}(t,T) = \langle \left(\phi(0,t) - \phi(0,0) \right)^2\rangle_T$
is the symmetric correlator of the $g=V=0$ model (i.e.\ Eq.~\ref{eq:ham}) at temperature $T$.
Since the symmetric correlator has a logarithmic divergence, we implement a UV cutoff $a^{-1}$ in the frequency. The correction to conductance then assumes the form
\begin{equation}
  \label{eq:conductance-correction}
  \Delta G(T) = - \frac{\pi e^2}{4 h}  g^2  a^2 {K'}^4 \left( \frac{a}{d} \right)^{\alpha} f(T d)
\end{equation}
where $f$ is a scaling function which depends only on the dimensionless parameter $Td$, and  %
$\alpha = 2K(0)-2$ is exponent of the UV divergence for a finite-length junction.

The temperature dependence of the resulting conductance correction $\Delta G$, for the piecewise-exponential profile of the Luttinger parameter $K(x)$, is shown in Fig.~\ref{fig:perturbative_temperature_conductance}. These results demonstrate that in the low-temperature limit--- but still above where the backscattering becomes large--- the suppression of the conductance matches the known power law $\Delta G\sim T^{-1}$~\cite{chklovskii1998consequences,chamon1997distinct,fendley1994exact}.
In the high-temperature limit,
  \begin{equation}
    \label{eq:high-temp-Delta-G}
    \Delta G(T) = - C \frac{e^2}{h} (g a^{K(0)})^2 \, T^{2K(0)-2}
    \end{equation} where $C$ depends only on $K'$ and $K(0)$ (see SM~\cite{supplemental} for exact form).
The scaling exponent corresponds to the Luttinger parameter at the impurity, recovering the result from Ref.~\onlinecite{chklovskii1998consequences}.
The crossover point between these two scaling exponents gives an experimentally-feasible way to determine the energy scale corresponding to the junction width $d$, provided that the difference between $K'$ and $K(0)$ is resolvable.

Additionally, the crossover in scaling exponent can quantitatively relate the microscopic impurity to the BSG energy scale.
The physics of the BSG model, which describes the low temperature regime, is characterized completely by a single parameter $\EB$ (cf.\ $T_0 = 2 \EB$ in Ref.~\onlinecite{cohen2023universal}).
For temperatures lower than the crossover point but pertubatively close to the QHT limit, the conductance correction follows the exact BSG value $\Delta G(T) = -(e^2/h) (\pi \EB / 8T)$~\cite{fendley1994exact,cohen2023universal}.
For the weak backscattering case $\EB \ll 1/d$, one can compare this high temperature BSG conductance to Eq.~\ref{eq:conductance-correction} in order to relate $\EB$ to %
the cutoff-invariant impurity strength  $g a^{K(0)}$ (from the scaling exponent of $g$ in Eq.~\ref{eq:ham}) as well as $d$. 
These microscopic parameters relate to the ac and dc conductances in Eq.~\ref{eq:full-harmonic-conductance} and \ref{eq:high-temp-Delta-G} or the full temperature dependence of the dc conductance Eq.~\ref{eq:conductance-correction} (see SM~\cite{supplemental} for details).

\section{Numerical model of ac conductance}%
Extending the results beyond weak backscattering requires numerical simulation.
Such a nonperturbative treatment is necessary when $\EB \gg 1/d$ and the crossover in exponents is not visible, as is the case in Ref.~\onlinecite{cohen2023universal}.
For this purpose, we approximate the model in Eq.~\ref{eq:ham} as an interacting fermion lattice model with next-nearest neighbor interactions~\cite{giamarchi2003quantum},
described by a Hamiltonian
\begin{multline}
\label{eq:numerical-ham}
    H=\sum_j \left(t_j \hat T_{j}-\mu_j n_j + U_j n_j n_{j+1}+U'_j n_j n_{j+2}\right) + u n_0
\end{multline}
where $\hat T_j = c^\dagger_j c_{j+1} + c^\dagger_{j+1} c_{j}$ and the spatially dependent parameters correspond to $K=1$ and $K=1/3$ Luttinger liquids with a short, smooth interface.\footnote{The next nearest neighbor interactions are helpful in achieving the $K=1/3$ state without a gap opening.}
For the $K=1$ side we use
$t_n = 0.308$, 
$\mu_n = 0.360$,
$U=0$ and
$U'_n=0$
while for the $K=1/3$ side we use
$t_n = 0.122$, 
$\mu_n = 1.22$, 
$U=0.293$ and 
$U'_n=0.439$.
These parameters were determined by the variational uniform matrix product state (VUMPS) algorithm~\cite{vumps} to realize the correct Luttinger parameter while keeping the velocity and density approximately constant.
The interface between the two sides consists of a smooth transition over three lattice sites in each of the parameters.

Since the long-wavelength description of this lattice model is the Luttinger liquid of Eq.~\ref{eq:ham}, we expect that $\EB$ should be a function of $g a^{K(0)}$ and $d$ in agreement with the perturbative argument.
  Since $g a^{K(0)}$ is cutoff-invariant under RG, we can arbitrarily choose $a$ to be the lattice constant (taken to be $1$) such that $\EB$ is only a function of $g$. 
  From the sliding CDW picture, $g$ is proportional to the impurity strength $u$.

\begin{figure}[t]
\centering
\includegraphics[width=\columnwidth]{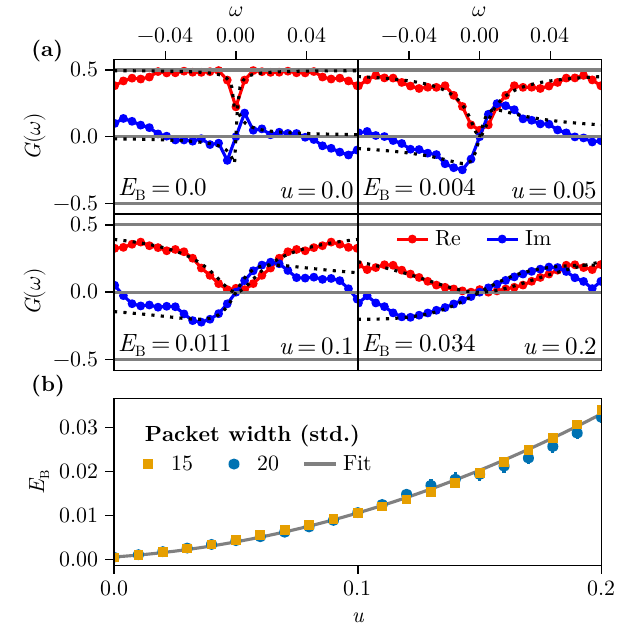}
\caption{\label{fig:ac_conductance} \textbf{(a)} ac conductivity for various impurity strengths $u$ and packet width $15$ demonstrating near $G=e^2/2h$ for $u=0$. The  data is fit to Eq.~\ref{eq:majorana_conductance} to calculate barrier energy $\EB$, with fits shown as dotted lines. \textbf{(b)} Fit value of $\EB$ as a function of impurity strength $u$, demonstrating a quadratic relationship. Data from the narrower packet is fit to $\EB=A(u+u_0)^2$, which yields $A=0.55$ and $u_0=0.04$, the latter of which represents the intrinsic backscattering of the Luttinger parameter crossover. 
}
\end{figure}%

We probe the ac conductance of this junction using a Gaussian chiral wavepacket, generated by a local quench in the chemical potential $\mu_n$ and a gauge field $a_n$ such that $t_n \mapsto t_n e^{ia_n}$~\cite{ganahl2012observation}.
We use the density matrix renormalization group~\cite{white1992,white1993} to solve for the ground state of this quenched Hamiltonian and then use the time-evolving block decimation algorithm~\cite{vidal2003,vidal2004} to evolve the packet in real time with a fourth-order Trotter decomposition~\cite{barthel2020}. 
The propagation of these packets is nearly dissipationless, since the packet is perturbatively small ($\rho \lesssim 10^{-3}$)
and propagates as a soliton,
as shown in Fig.~\ref{fig:trajectory}.
We see that in the dc limit, the backscattering energy is relevant and drives the conductance to zero while at higher frequencies, we recover the QHT limit.
This phenomenon manifests in Fig.~\ref{fig:trajectory} as the long wake of normal reflection following the sign-flipped charge packet.
Comparing the initial and transmitted packet in the frequency domain gives the ac conductance, plotted in Fig.~\ref{fig:ac_conductance}a.
Note that due to the discrete geometry and the finite bandwidth of the Gaussian wavepacket, there is a trade-off between frequency range and resolution, and therefore two different packet widths are used.
Numerical details can be found in the SM~\cite{supplemental}.

\section{Boundary sine-Gordon model}
In this low-energy limit where the wavelength is much longer than the junction width, we can understand the numerical results by mapping the non-chiral Luttinger liquid in Eq.~\ref{eq:ham} to the boundary sine-Gordon model (BSG)~\cite{fendley1994exact} which is exactly solvable using refermionization.
By folding the fields at $x<0$ using a reflection $x\to -x$, we obtain a theory of two fields on the positive half-line which in the short junction limit have constant Luttinger parameter.
The smooth junction then becomes a pair of boundary conditions at $x=0$.
By performing a linear transformation on this pair of fields, we obtain two decoupled fields $\xi$ and $\bar\xi$ which have boundary conditions $\partial_x \xi(0) = 0$ and $\bar\xi(0) = 0$.
Since $\bar\xi$ vanishes at $x=0$, the original field at the boundary can be written purely in terms of $\xi(0)$ as 
$\phi(0)=\Gamma \xi(0)$ where $\Gamma \equiv (K_\L^{-1}+K_\R^{-1})^{-1/2}$.
The $\bar\xi$ equation has no boundary term and cannot affect the current, so we focus on the $\xi$ part of the Hamiltonian which takes a BSG form:
\begin{align}
    \label{eq:ham2}
     H_\xi=&\int_0^{\infty} \frac{\mathrm dx}{2\pi}\, \left[\eta^2+(\partial_x\xi)^2\right] %
     -\tilde{g} \cos{[2\Gamma \xi(0)]}.
\end{align}
where $\eta$ is the canonical momentum of $\xi$
and $\tilde g$ is the renormalized value of $g$ after the junction width $d$ is flowed to the UV cutoff $a$.

The above Hamiltonian matches the BSG Hamiltonian in previous works~\cite{sandler1998andreev,fendley1995exact} with the effective Luttinger parameter $K'=2\Gamma^2$,
giving the scaling exponent for the BSG impurity $\tilde g$ which matches the non-chiral Luttinger liquid exponent in the $d\rightarrow 0$ limit~\cite{safi1999b}.
When written in terms of $\xi(0)$, the voltage perturbation $\pi^{-1} V(t)\phi(0)$ to the Luttinger liquid becomes an effective voltage  $\Veff=\Gamma V$ acting on $\xi$ and a current $j = \Gamma \pi^{-1}  \eta(0)$. %

In order to refermionize and solve the BSG Hamiltonian, we temporarily represent the field as a single chiral mode on the entire real line, written as 
$\chiral\xi = (1/2)\left[ \xi(|x|) + \int_0^x dx' \eta(|x'|)\right]$.
Following Refs.~\onlinecite{shankar1995bosonization,von1998bosonization},
we define the chiral fermion field 
\begin{equation}
\label{eq:refermionization}
\chiral\psi(x) = \frac{\alpha}{\sqrt{2\pi a}} \, e^{-2i\chiral\xi(x)},
\end{equation}
where $\alpha=\exp[ i\pi(\chiral\xi(0)-\chiral\xi(\infty))]$ 
is the fermion parity of the refermionized model and $a$ is the UV cutoff.
The boundary term, for the specific case of $\Gamma=1/2$ corresponding to Luttinger parameters $1/3$ and $1$, can be written in terms of this chiral fermion operator as
\begin{align}\label{eq:cosF}
    \cos{[2\chiral\xi(0)]}= \sqrt{\frac{\pi a }{2}} \, \alpha \left[ \chiral{\psi}(0)+ \chiral{\psi}^\dagger (0)\right].
\end{align}
Note that this term is quadratic in fermion operators due to the Majorana operator $\alpha$. 

Eq.~\ref{eq:cosF} represents two transmission processes mediated by the local Majorana operator $\alpha$: one which  preserves charge and one which flips charge. When refolded onto the half-line of Eq.~\ref{eq:ham2}, these become normal and Andreev reflection processes at the boundary with amplitudes
\begin{align}\label{eq:rAndreev}
& \rN=-\frac{\EB}{i\omega+\EB},\quad\,  \rA=\frac{i\omega}{i\omega+\EB}
\end{align}
where $\EB=\tilde g^2\pi a/4$ is the resultant energy scale.
At low energies $ |\omega|\ll |\EB|$ the normal reflection dominates while high-energy fermions with 
$|\omega|\gg |\EB|$ experience Andreev reflection.

This quantitative result is consistent with previous works solving Eq.~\ref{eq:ham2} using a Kramers-Wannier duality~\cite{guinea1985dynamics,sandler1999noise} and conformal field theory techniques~\cite{fendley1994exact,ghoshal1994,ameduri1995boundary}.
The voltage and temperature dependence resulting from these scattering amplitudes~\cite{chamon1997distinct,wimmer2011quantum} agrees with the BSG predictions in the tunneling interpretation, though the sliding charge density model used here is microscopically better motivated.
Additionally, the results are in 
good agreement with recent experiment~\cite{cohen2023universal}.

To connect with the numerical results, we calculate the ac conductance associated with these amplitudes using the formalism of dynamic conductance~\cite{buttiker1993dynamic}:
\begin{align}
    G(\omega)%
    &=\frac{e^2}{2h} \left[ 1 - \frac{\EB}{i\omega} \, \mathrm{Ln} \left(1 + \frac {i\omega} {\EB}\right)\right],\label{eq:majorana_conductance}
\end{align}
where $\mathrm{Ln}$ is the principal value of the complex natural logarithm.
The real part of the ac conductance matches the results from the analysis of a  
$K=1/2$ BSG model~\cite{zamoum2012}  and vanishes as $\omega\rightarrow 0$, which is consistent with the irrelevance of the tunneling perturbation~\cite{kane1992transmission}. Furthermore, it 
 reaches the harmonic value of $e^2/2h$ as $\omega\rightarrow \infty$, which is consistent
 with Eq.~\ref{eq:full-harmonic-conductance} when $d^{-1}\gg\omega\gg \EB$.
We fit the ac conductances in Fig.~\ref{fig:ac_conductance}a with Eq.~\ref{eq:majorana_conductance} and find a close resemblance at low frequencies, where the ratio of Gaussians is well defined. Furthermore, we can deduce the values of $\EB$ from different values of the impurity strength $u$ as shown in Fig.~\ref{fig:ac_conductance}b, which are consistent with expectations  ~\cite{fukuyama1978dynamics} apart from a shift $u_0$ that likely arises from 
the spatial variation of $K(x)$.

\section{Sliding CDW model for the QPC}%
While the non-chiral Luttinger liquid Hamiltonian (Eq.~\ref{eq:ham}) implicitly assumes a QPC width on the order of the magnetic length, the QPC used in the experimental measurements~\cite{cohen2023universal} is much wider.
To resolve this discrepancy, we consider the effect of a smooth (on the scale of the magnetic length) disorder potential in the vicinity of the pinch-off of the QPC.
Such a random potential inevitably leads to localized states that fill the entire width when the gate potentials have completely pinched off the QPC.
Here, we assume that a small change in the gate potential from complete pinch-off allows a single percolating path of electrons that would appear at conductances slightly above pinch-off. 

The low density of electrons in this channel, which are strongly interacting, forms a sliding CDW with a spatially varying density interpolating between $\nu=1/3$ and $\nu=1$.
Such a state would be a ballistic Luttinger liquid consistent with the QHT behavior~\cite{chklovskii1998consequences} for a smooth and weak disorder potential.
This picture is consistent with Fig.~S3A in Ref.~\onlinecite{cohen2023universal} which shows a robust $G=e^2/2h$ region just above the pinch-off. 
At higher voltages, the clear QHT behavior is replaced by large fluctuations which indicate the presence of multiple percolating modes that are beyond the simple model considered here.
Multiple experimental techniques should be able to verify this effective QPC model of a wide disordered junction, including scanning acoustic tomography, scanning superconducting quantum interference devices (SQUIDs), and microwave impedance microscopy.

\section{Conclusion}
In this Letter we study the 
 consequences of a pinning potential in the CDW model and show that the transport matches the BSG model~\cite{chamon1997distinct} in the low temperature and frequency limit. 
We also study the high temperature regime and derive perturbative corrections to the QHT conductance from the pinning potential, which can be used to identify the subleading energy scale associated with the finite length of the junction and the microscopic impurity strength.
Then, a matrix product state calculation extends these results beyond the low frequency range at zero temperature, achieving both sign-flipping reflection in the QHT limit and zero conductance in the zero-frequency limit.
Comparing this calculation to the BSG leads to a quadratic relationship between the impurity strength $u$ and the backscattering energy $\EB$.
Combining these two methods gives a concrete experimental prediction: Measuring the impurity strength and junction width using the high-temperature perturbation method and inputting these values into the nonperturbative numerical method should give the same $\EB$ value observed from the low-temperature conductance.

The microscopic connection between experiment~\cite{cohen2023universal} and the BSG model described in this work could guide future experimental studies, providing predictions for the microwave conductivity~\cite{zamoum2012} as well as shot noise and dc conductivity~\cite{chamon1997distinct,sandler1998andreev,sandler1999noise,zamoum2012} at low temperature and voltage.
Through these observables, the QPC geometry can provide a window to the physics of the BSG model.
Such physics could be realized in anyon collider experiments~\cite{bartolomei2020} which are also amenable to the methods proposed here.
In particular, the flexibility of matrix product states allows one to simulate a wide range of junction widths, impurity profiles and fillings.
Recent work has explored QPC junctions with Moore--Read and Read--Rezayi states~\cite{ohashi2022,ohashi2026,ahmed2026}, which could serve as an extension of the methods proposed here.

\begin{acknowledgments}
We thank Bertrand Halperin, Michael Zaletel and Andrea Young for valuable discussion.
S.Y.-T.\@ thanks the Joint Quantum Institute at the University of Maryland for support through a JQI Graduate Fellowship.
J.S.\@ acknowledges support from the Joint Quantum Institute and the hospitality of the Aspen Center for Physics, which is supported by National Science Foundation grant PHY-2210452,  during the final 
stages of this work.
This work is also supported by the Laboratory for Physical Sciences through its continuous support of the Condensed Matter Theory Center at the University of Maryland.
\end{acknowledgments}%

\section{Data availability}%
The data that support the findings of this article are openly available~\cite{zenodo}.

\nocite{kane1994randomness,quadgk,vlijm2015quasi,muzykantskii1994quantum,de1994doubled}
\bibliography{bib.bib}

\end{document}


\title{Supplementary Material for ``Quantum Hall transformer in a quantum point contact over the full range of transmission''}

\author{Stuart Yi-Thomas}
\email[]{snthomas@umd.edu}
\affiliation{Condensed Matter Theory Center and Joint Quantum Institute, Department of Physics, University of Maryland, College Park, Maryland 20742-4111, USA}
\author{Jay D. Sau}
\affiliation{Condensed Matter Theory Center and Joint Quantum Institute, Department of Physics, University of Maryland, College Park, Maryland 20742-4111, USA}

\date{October 27, 2026}

\maketitle
\renewcommand\thefigure{S\arabic{figure}}    
\renewcommand{\theequation}{S\arabic{equation}}

\section{Perturbative method Details}
\subsection{Equation of motion}
This section contains a detailed derivation of the perturbative calculation leading to Eqs.~2-4 of the main text.
We begin with the full Luttinger liquid Hamiltonian (Eq.~1 of the main text) with a localized impurity and a voltage $V(t)$, applied at a site $x_0$.
We replace $\Pi$ with $\partial_x \theta$ such that $[\theta(x), \phi(x')] = (i\pi/2) \mathrm{sgn}(x-x')$.
For generality, we take the potential to be spatially dependent, given by $g(x)$, but assume that it is localized to the transition region. The Luttinger liquid Hamiltonian is
\begin{align}
    H=\int dx\, \frac{v}{2\pi}\left[ K(x) (\partial_x \theta)^2+K(x)^{-1}(\partial_x\phi)^2\right] + g(x) \cos 2\phi + \pi^{-1} V(t) \phi(x_0).
\end{align}
The corresponding Lagrangian is
\begin{align}
    L= \int dx\,  \frac 1 {2\pi K(x)} \left[ v^{-1} (\partial_t \phi)^2 -  v (\partial_x\phi)^2\right] - g(x) \cos 2\phi - \pi^{-1}V(t) \phi(x_0).
\end{align}
The equations of motion are
\begin{align}
\dot \phi = i [H, \phi] &=  -v K(x) \, \partial_x \theta \\
\dot \theta = i [H, \theta] &= -\frac{v}{K(x)} \, \partial_x \phi + i \int dx' \, g(x') [\cos2\phi(x'), \theta(x)] - \frac{V(t)}{2}\mathrm{sgn}(x-x_0)\\
&= -\frac{v}{K(x)} \, \partial_x \phi - \pi \int dx' \,\mathrm{sgn}(x-x') g(x') \sin2\phi(x') -  \frac{V(t)}{2}\mathrm{sgn}(x-x_0)
\end{align}
which combine to give
\begin{equation}
\frac{1}{\pi K(x)} \left( \ddot\phi - \partial_x^2 \phi + \frac{K'(x)}{K(x)}  \partial_x \phi \right) = 2 g(x) \sin2\phi + \pi^{-1} V(t) \delta(x-x_0)
\end{equation}
where we set $v=1$ for simplicity.
The left-hand side then becomes the homogeneous differential operator of the Gaussian theory.

When $K(x)$ is constant, the Luttinger parameter cancels out and the equation has plane wave solutions. The equation is also exactly solvable when $K'(x) / K(x)$ is constant. Specifically for $K(x)\propto e^{2 A x}$,
\begin{equation}
  \label{eq:pde}
v^{-2} \ddot\phi = \partial_x^2 \phi - 2 A \partial_x \phi
\end{equation}
which gives solutions $\phi \propto e^{r_\pm x}$ where
\begin{align}
r_\pm = A \pm \sqrt{A^2 - (\omega/v)^2}.
\end{align}
for a frequency $\omega$.
We therefore model the junction as a piecewise exponential function with a transition region of length $d$, constrained to be continuous:
\begin{equation}
\label{eq:piecewise}
K(x) = \begin{cases}
K_- & x<-d/2 \\
\sqrt{K_\L K_\R} \, \left(\frac{K_+}{K_-}\right)^{x/d} & -d/2\leq x<d/2 \\
K_+ & x\geq d/2 \\
\end{cases}.
\end{equation}
Therefore $A=(1/2d)\ln(K_-/K_+)$.
When $K_-/K_+$ is close to $1$, the exponential portion is approximately linear. 

\subsection{Retarded Green's function}
We now compute the retarded Green's function $\mathcal G$ of the homogeneous problem.
This can be constructed using the homogeneous solutions and matching the derivatives and function values at the boundaries. Since we are interested first in the retarded Green's function, we solve for the plane-wave solutions with outgoing modes at $x=\pm \infty$.
Note that the factor of $1/\pi K(x)$ is included in the homogeneous differential operator. Matching boundary conditions, we obtain

\begin{align}
  \label{eq:greens-function}
  \mathcal G(x, x'; \omega) = \frac{i \pi \sqrt{K(x) K(x')}}{2 (\omega - i \epsilon)} \frac{\, g_-\left[-\frac{d}{2} - \min(x,x')\right] \, g_+\left[\max(x,x')-\frac{d}{2}\right]}{\cos \tilde\omega d - (i \omega / \tilde\omega)\sin \tilde\omega d} 
\end{align}
where
\begin{align}
  g_\pm(x) &= \begin{cases} \cos \tilde \omega x + \frac{i \omega \mp A}{\tilde \omega} \sin \tilde \omega x & x < 0 \\
    e^{i \omega x} & \text{otherwise}.
  \end{cases}
\end{align}

\subsection{Secord order expansion in backscattering}
We can introduce the voltage term by shifting the homogeneous field, giving an effective driven field
\begin{align}
  \phi_0'(x,t) &= \phi_0(x,t) + f(t) \hspace{1cm} \text{where} \hspace{1cm} f(x,t) = \pi^{-1} \int dt'\, \mathcal G(x,x_0; t-t')  V(t').
\end{align}
The total solution is then given by the self-consistent equation
\begin{align}
\phi(x,t) = \phi_0'(x,t) + 2\int dx'dt' \, g(x') \mathcal G(x,x';t-t') \sin2\phi(x',t').
\end{align}
A second-order Born approximation gives
\begin{align}
\phi(x,t) &= \phi_0'(x,t) + 2\int dx'dt' \, g(x') \mathcal G(x,x';t-t') \sin\left[ 2\phi_0'(x',t') + 4\int dx''dt'' \, g(x'') \mathcal G(x',x'';t'-t'') \sin2\phi_0'(x'',t'')\right]
\end{align}
which to second order in $g$ and first order in $V$ becomes
\begin{multline}
\phi(x,t) = \phi_0'(x,t) + 4 \int dx'dx''dt'dt'' \, g(x') g(x'')  \mathcal G(x,x';t-t') \mathcal G(x',x'';t'-t'') \\ \times \left[2f(t'') - 2f(t')\right]  \cos\left[2\phi_0(x'',t'') - 2\phi_0(x',t')\right],
\end{multline}
ignoring terms that are not symmetric under a shift in $\phi_0$ since their expectation value disappears.
The expectation value of $\phi$ can be calculated using Wick's theorem at a finite temperature $T$:
\begin{align}
\langle \phi(x,t) \rangle_T &= f(x,t) + 4 \int dx'dx''dt'dt'' \, g(x') g(x'')  \mathcal G(x,x';t-t') \mathcal G(x',x'';t'-t'')  \nonumber \\
&\hspace{4cm} \times \left[2 f(x'',t'') - 2 f(x',t') \right] \exp\left[-2 \mathcal G_\text{S}(x',x'';t'-t'',T) \right],
\end{align}
where
\begin{align}
\mathcal G_\text{S}(x,x';t,T)= \left\langle \left[ \phi_0(x,t) - \phi_0(x',0) \right]^2 \right\rangle_T
\end{align}
is the temperature-dependent symmetric Green's function.
\subsection{Symmetric Green's function}
The two-point function $G_\text{S}$ can be expanded as
\begin{align}
  \mathcal G_\text{S}(x,x';t,T) &= \left\langle \left[\phi(x,t) - \phi(x',0)\right]^2\right\rangle \\
                 &= \langle \phi(x,t) \phi(x,t) \rangle + \langle \phi(x',0) \phi(x',0) \rangle - \langle \phi(x,t) \phi(x',0) \rangle - \langle \phi(x',0) \phi(x,t) \rangle \\
                 &= \mathcal G^>(x,x;0) + \mathcal G^<(x',x';0) - \mathcal G^>(x,x';t) - \mathcal G^<(x,x';t).
\end{align}
The Wightman functions for the first two terms are arbitrary since they are equal at $t=0$.
Both $G^{>}$ and $G^{<}$ can be calculated from the spectral function in the transition region
\begin{align}
  \rho(x,x';\omega) &= \mathcal G^> - \mathcal G^< \\
  &= 2 \mathrm{Im} \mathcal G(x,x'; \omega) \\
            &= \frac{ \pi \sqrt{K(x) K(x')} }{\omega} \frac{\omega ^2 \cos [\tilde{\omega}  (x-x')]-A \cos (\tilde{\omega}  d) \left(A \cos [\tilde{\omega}  (x+x')]+\tilde{\omega}  \sin [\tilde{\omega} (x+x')]\right)}{\omega ^2-A^2 \cos ^2(\tilde{\omega}  d)} . \label{eq:spectral-function} %
\end{align}
From the Kubo-Martin-Schwinger condition we have $\mathcal G^<(x,x';\omega) = e^{\beta \omega} \mathcal G^>(x,x';\omega)$ and thus
\begin{align}
  \mathcal G^>(x,x';\omega) &= [1+n(\omega)] \rho(x,x';\omega) \\ 
  \mathcal G^<(x,x';\omega) &= n(\omega) \rho(x,x';\omega)
\end{align}
where $n(\omega) = 1/(e^{\beta \omega} - 1)$ is the Bose-Einstein distribution.
This gives a final expression for $G_\text{S}$, noting that $\rho(x,x';\omega) = \rho(x',x;\omega)$ and $\int d\omega \rho(x,x;\omega)=0$:
\begin{align}
  \mathcal G_\text{S}(x,x';t,T) %
  &= \int \frac{d\omega}{2\pi} \coth \frac{\omega}{2T} \left[ \frac{1}{2} \rho(x,x;\omega) + \frac{1}{2} \rho(x',x';\omega) - \cos \omega t \;   \rho(x,x';\omega) \right].
\end{align}
\begin{newsection}
  When $x=x'=0$, this function becomes
\begin{align}
  \mathcal G_\text{S}(0,0;t,T) &= \int \frac{d\omega}{2\pi} \coth \frac{\omega}{2T} \left( 1 - \cos \omega t \right)   \rho(0,0;\omega) \\
   &= \pi K(0) \int \frac{d\omega}{2\pi} \coth \frac{\omega}{2T} \frac{1 - \cos \omega t }{\omega} \frac{\omega^2 - A^2 \cos \tilde \omega d}{\omega^2 - A^2 \cos^2 \tilde \omega d}.
\end{align}
\end{newsection}
These expressions have a UV divergence in the frequency which scales logarithmically in the distance cutoff $a$, coming from the $1/\omega$ prefactor in Eq.~\ref{eq:spectral-function}. The logarithmic divergence has a coefficient $K(0)=\sqrt{K_\L K_\R}$ such that
\begin{equation}
  \label{eq:a-scaling}
  \mathcal G_\text{S}(x,x';t,T) - K(0) \ln \left(\frac{a}{d}\right)
\end{equation}
converges in vanishing $a$. (The denominator $d$ plays the role of an arbitrary length scale such that the argument of the logarithm is dimensionless.) If we instead take the $d\to 0$ limit first (that is $d\ll a$), the prefactor $K(0)$ becomes $K' = 2 / (K_\L^{-1} + K_\R^{-1})$.

\subsection{Conductance measurement}
If the driving voltage has a frequency $\omega_0$, we can assume that $f(x,t)$ oscillates in time with frequency $\omega_0$.
This allows us to simplify the algebra by writing the Fourier transform of $f(x,t)$ as $f (x) \delta(\omega - \omega_0)$ and taking the real part of $\langle \phi(x,\omega) \rangle_T$:
The expression using convolutions allows us to easily write the field expectation value in frequency space.
The value is only nonzero when $\omega=\pm \omega_0$. We take $\omega = \omega_0$ without loss of generality since the opposite is attainable via complex conjugation.
\begin{align}
\langle \phi(x,\omega_0) \rangle_T &= f(x) + 8 \int dx'dx''\, g(x') g(x'')  \mathcal G(x,x';\omega_0) \int dt \, \mathcal G(x',x'';t'-t'') e^{-2\mathcal G_\text{S}(x',x'';t'-t'',T)} \left[ e^{i\omega_0 t} f(x'') - f(x') \right].
\end{align}
Substituting $f(x) = \pi^{-1} V \mathcal G(x,x_0; \omega_0)$, we obtain
\begin{multline}
\frac{\pi \langle \phi(x,\omega_0) \rangle_T}{V} = \mathcal G(x,x_0;\omega_0) + 8 \int dx'dx''\, g(x') g(x'')  \mathcal G(x,x';\omega_0) \\ \times \int dt \, \mathcal G(x',x'';t'-t'') e^{-2\mathcal G_\text{S}(x',x'';t'-t'',T)} \left[ e^{i \omega_0 t} \mathcal G(x'',x_0; \omega_0) - \mathcal G(x',x_0;\omega_0) \right].
\end{multline}

The current is related to the time derivative of the expectation value of $\phi$ such that $I= \pi^{-1} \langle   \partial_t \phi\rangle$. Therefore the ac conductance is
\begin{equation}
  G(\omega,T) = \frac {I(\omega)} {V(\omega)} = \frac{\omega \langle \phi(x,\omega)\rangle_T}{i\pi V} =  \frac {e^2}{h} \frac{2 \omega }{i\pi} \left( \frac{\pi \langle \phi(x,\omega) \rangle_T}{V} \right),
\end{equation}
noting that $e^2/h=1/2\pi$ in natural units. 
Substituting the perturbative expression gives $G(\omega,T) = G_0(\omega)- \Delta G(\omega,T)$ where $G_0(\omega)$ is the harmonic conductance and $\Delta G(\omega,T)$ is the second-order correction in $g$.

The harmonic conductance is proportional to the Green's function from $x_0=-d/2$ to $x=d/2$:
\begin{align}
  G_0(\omega) &= \frac{e^2}{h} \frac{2\omega}{i\pi} \mathcal G(-d/2,d/2;\omega)
\end{align}
which by Eq.~\ref{eq:greens-function} is
\begin{align}
  G_0(\omega) &= \frac{e^2}{h} \frac{\sqrt{K_\L K_\R}}{\cos \tilde\omega d - (i \omega /\tilde \omega) \sin \tilde\omega d},
\end{align}
leading to Eq.~2 in the main text.

The correction to conductance takes the form
\begin{align}
  \label{eq:linear-response-eq}
 \Delta G(\omega,T) = \frac{e^2}{h} \frac{16\omega}{i \pi} \int dx'dx''\, g(x') g(x'')  \mathcal G(d,x';\omega)  \int dt \, t \; \mathcal G(x',x'';t ) e^{-2\mathcal G_\text{S}(x',x'';t, T)} \left[ e^{i \omega t} \mathcal G(x'',0; \omega) - \mathcal G(x',0;\omega) \right].
\end{align}

To calculate the dc conductance at finite temperature, we take the limit as frequency goes to zero. In this limit, the retarded Green's function takes the spatially-independent form 
\begin{align}
  \label{eq:dc-limit-G}
  \mathcal G(x,x';\omega) = \frac{i \pi K'}{2(\omega - i\epsilon)}
\end{align}
where $K'=2/(K_-^{-1} + K_+^{-1})$ is the effective Luttinger parameter. Then
\begin{align}
  \label{eq:linear-response-eq-intt}
  \Delta G(T) = -\frac{4e^2}{h} \, \pi {K'}^2 \int dx'dx''\, g(x') g(x'') \int dt \, t \; \mathcal G(x',x'';t ) e^{-2\mathcal G_\text{S}(x',x'';t, T)} .
\end{align}

For simplicity, we now consider $g(x) = g \delta(x)$.
Rescaling $t$ to $\tau d$ as
\begin{align}
  \Delta G(T) = -\frac{4e^2}{h} \, \pi {K'}^2 d^2 g^2 \int d\tau \, \tau \; \mathcal G(0,0;\tau d ) e^{-2\mathcal G_\text{S}(0,0;\tau d, T)}
\end{align}
leads to an integrand which is only dependent on the dimensionless parameters $Td$ and $a/d$, since $d^{-1} \rho(\xi d,\xi' d;\nu/d)$ is independent of $d$ for dimensionless parameters $\xi$, $\xi'$ and $\nu$.
To isolate the UV divergence, we pull out the logarithmic term such that the exponent matches the nondivergent expression in Eq.~\ref{eq:a-scaling}:
\begin{align}
  \Delta G(T) = -\frac{4\pi e^2}{h} \, g^2 a^2 {K'}^2  \left(\frac{a}{d}\right)^{2K(0)-2} \int d\tau \, \tau \; \mathcal G(0,0;\tau d ) \, e^{-2\left[ \mathcal G_\text{S}(0,0;\tau d, T) - K(0) \ln (a/d)\right]}.
\end{align}
When $a$ is very small compared to $d$ and $1/T$, the integral factor depends only on the parameter $Td$.
\begin{newsection}
\subsection{Long junction}
In the long junction case, i.e.\ $d \gg 1/T$, the impurity exists in a locally homogeneous region with Luttinger parameter $K(0)$.
Therefore the spectral function becomes
\begin{align}
  \rho(x,x';\omega) = \frac{\pi K(0)}{\omega}
\end{align}
and the exponential becomes
\begin{align}
  e^{-2\left[ \mathcal G_\text{S}(0,0;\tau d, T) - K(0) \ln (a/d)\right]} =  e^{- 2 K(0) \gamma} \left[ \frac{\pi T d}{\sinh(\pi T t)}\right]^{2 K(0) - 2} 
\end{align}
where $\gamma$ is Euler's constant. Therefore the conductance correction is
\begin{align}
  \label{eq:long-junction-correction}
  \Delta G(T) = - C \frac{e^2}{h} (g a^{K(0)})^2  \, T ^{2K(0) - 2}
\end{align}
where
\begin{equation}
  \label{eq:I-constant}
  C = 2 \pi^2 {K'}^3 e^{-2 K(0) \gamma} \int^\infty_0 ds \, s \left[\frac{\pi}{\sinh(\pi s)}\right]^{2K(0)}.
\end{equation}
In order to isolate the dimensionless impurity strength $g a$, one can extract $K(0)$ from the exponent with temperature and calculate $g a$ from the prefactor.

\end{newsection}

\subsection{Short junction}
In the short junction limit, when $d \ll a$, the divergence coefficient is $K'$ and 
\begin{align}
  \Delta G(\omega,T) = -\frac{4 \pi e^2}{h} \, g^2 a^2 \, {K'}^2  \left(\frac{a}{d}\right)^{2K'-2} \int d\tau \, \tau \; \mathcal G(0,0;\tau d ) \, e^{-2\left[ \mathcal G_\text{S}(0,0;\tau d, T) - K' \ln (a/d)\right]}.
\end{align}
In this regime, the propagators simplify considerably.
The parameter $A$ is much larger than $\omega > 1/a$ which is equivalent to the low frequency limit, where the retarded propagator is given in Eq.~\ref{eq:dc-limit-G} and the spectral function is
\begin{equation}
  \rho(x,x';\omega) = \frac{\pi K'}{\omega}
\end{equation}
in the transition region.
The symmetric Green's function is then
\begin{align}
  \mathcal G_\text{S}(x,x';\tau d,T) - K' \ln\left(\frac{a}{d}\right) &=
                                  K' \int_0^{d/a} d\nu \coth \frac{\nu}{2Td} \frac{1-\cos (\nu \tau) }{\nu} - K' \ln\left(\frac{a}{d}\right) \\
  &= K' \left[ \ln \frac{\sinh(\pi T d \tau)}{\pi T d \tau} + \gamma + \ln\tau \right],
\end{align}
ignoring terms which disappear for small $a$,
leading to 

\begin{align}
  e^{-2\left[\mathcal G_\text{S}(x,x';\tau d,T) - K' \ln\left(\frac{a}{d}\right)\right] }
  &= \left(\frac{\pi T d}{\sinh(\pi T d\tau)}\right)^{2 K'} e^{-2K'\gamma}
\end{align}
The time-domain retarded propagator in the dc limit becomes
\begin{equation}
  \mathcal G(x,x';t) = \frac{\pi K'}{2} \Theta(t)
\end{equation}
such that $\Delta G$ is
\begin{align}
  \Delta G(\omega,T) = -\frac{4\pi e^2}{h} \, g^2 a^2 \, {K'}^2  \left(\frac{a}{d}\right)^{2K'-2} \frac{\pi K'}{2} e^{-2K'\gamma} \int_0^\infty d\tau \, \tau \left(\frac{\pi T d}{\sinh(\pi T d \tau)}\right)^{2 K'},
\end{align}
which is solvable using beta and digamma functions. It has a scaling form
\begin{equation}
  \Delta G \sim - T^{2K'-2},
\end{equation}
matching Ref.~\onlinecite{kane1994randomness}.
In the case that $K'=1/2$, the integral becomes $\pi / (4Td)$, giving a conductance correction
\begin{align}
  \label{eq:my-short-junction-correction}
  \Delta G(\omega,T) &= -\frac{e^2}{h} \frac{\pi^3}{16} e^{-\gamma} \frac{g^2 a}{T}.
\end{align}
\begin{newsection}
From Refs.~\onlinecite{cohen2023universal,fendley1994exact}, the BSG conductance correction is 
\begin{align}
  \label{eq:short-junction-correction}
  \Delta G(T) &= -\frac{e^2}{h} \frac{\pi}{8} \frac{\EB}{T}
\end{align}
which matches the relationship $\EB \sim g^2 a$.
Such a relationship is expected when $d \ll a$ and the scaling exponent of $g$ is $1 - K' = 1/2$, however in the experiment the junction width is relevant at high temperatures~\cite{cohen2023universal}.
\end{newsection}

\begin{newsection}
\subsection{Calculating $\EB$ from microscopic parameters}
From Eq.~\ref{eq:conductance-correction} and Fig.~2 in the main text, the scaling exponent of the conductance correction is
\begin{equation}
  \frac{d \log|\Delta G|}{d \log T} = Td \frac{f'(Td)}{f(Td)}
\end{equation}
and depends only on $Td$, interpolating between $2K'-2$ and $2K(0) - 2$. Therefore
\begin{equation}
  \frac{\Delta G(T_\text{H})}{\Delta G(T_\text{L})} = \exp \left[ \int^{T_\text{H} d}_{T_\text{L} d} ds \, \frac{f'(s)}{f(s)} \right]
\end{equation}
for temperatures $T_\text{H}$ and $T_\text{L}$ that are above and below the crossover respectively. Using the BSG expression in Eq.~\ref{eq:short-junction-correction} and the derived high-temperature expression in Eq.~\ref{eq:long-junction-correction} leads to an expression for $\EB$:
\begin{equation}
  \EB = \frac{8 C}{\pi} (g a^{K(0)})^2 \, T_\text{H}^{2K(0)-2} \, T_\text{L} \, \exp \left[ -\int^{T_\text{H} d}_{T_\text{L} d} ds \, \frac{f'(s)}{f(s)} \right].
\end{equation}
Provided that $T_\text{H}$ and $T_\text{L}$ are far from the crossover point, the result should be independent of particular choice.
Since $d$ can be computed from the position of the crossover (in units of $1/\si{K}$), this expression gives a direct mapping between the cutoff-invariant impurity strength $g a^{K(0)}$ and the energy scale $\EB$.
\end{newsection}

\begin{newsection}
\subsection{High-temperature exponents in experimental conductance data}
\begin{figure}[h]
\centering
\includegraphics[width=0.6\columnwidth]{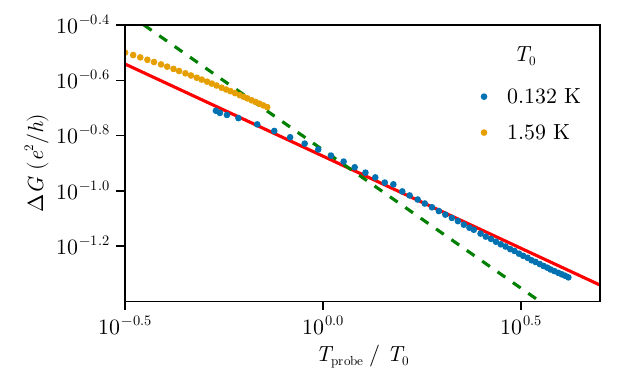}
\caption{\label{fig:exp_data_exponent}
  High-temperature scaling of experimental data from Ref.~\onlinecite{cohen2023universal}.
  The red solid line is the power law with exponent $-2/3$, corresponding to $K = 2/3 \approx K(x=0)$ and indicating that $d \gg 1/T$.
  The green dashed line corresponds to the exponent $-1$ predicted from the BSG.
  }
  
\end{figure}
By plotting the high-temperature data from Fig.~3 of Ref.~\onlinecite{cohen2023universal}, we extract the scaling exponent and find that it matches the long-junction exponent [$2 K(0) - 2 \approx - 2/3$] past the validity of the perturbative regime.
This observation is demonstrated in Fig.~\ref{fig:exp_data_exponent}.

Since there is no observed $-1$ power law, the conductance diverges from the BSG prediction at temperatures lower than the BSG energy scale $\EB$. Therefore we conclude that $1/d \ll \EB$ in this experiment.
An experiment may be able to observe the crossover in Fig.~2 of the main text if the impurity strength $\EB$ is reduced to smaller values, extending the pertubative regime to lower temperatures.

\end{newsection}

\subsection{Numerical implementation}
To numerically calculate $\Delta G(T)$ for arbitrary width $d$, we numerically evaluate this integral using Gauss-Kronrod quadrature~\cite{quadgk}.
We can also improve the numerical efficiency by subtracting off the asymptotic part for large frequencies and evaluate this analytically.

\section{Matrix Product State Details}
The $K=1/3$ Luttinger liquid poses a challenge to simulate numerically since there is no explicit form for the parameters in Eq.~\ref*{eq:numerical-ham} of the main text which lead to such a scaling coefficient. To calculate the corresponding parameters numerically, we measure the Drude weight and the compressibility and relate them back to the Luttinger parameters as $D=Kv/2\pi$ and $\chi=K/\pi v$. We use the VUMPS algorithm \cite{vumps} to determine these quantities for infinite systems to remove finite-size effects, particularly in calculating the compressibility. The Luttinger parameter is first set by optimizing the ground state $\psi$ over $K^2(\psi)=2\pi^2 D(\psi) \chi(\psi)$. Then, the velocity is tuned by rescaling $t_n$ and $\mu_n$. We use a large field to avoid the gapped phase near $\mu_n=0$ when $U_n>0$. We find that 
$t_n=0.122$, 
$U=0.293$, 
$U'_n=0.439$ and 
$\mu_n=1.22$ 
gives parameters of $v=0.5$\footnote{We choose $v=0.5$ instead of $v=1$ to access smaller frequencies without increasing the wavelength.} and $K=0.332$. To match the magnetization and velocity on the $K=1$ side, we set the hopping to 
$t_n = 0.308$ and the chemical potential to 
$\mu_n =(4t_n^2-v^2)^{1/2}= 0.360$.

We represent the interface between these two phases as a slow variation of $K(x)$ relative to the lattice spacing, which corresponds to slow variations in the lattice parameters. The Hamiltonian in Eq.~\ref*{eq:numerical-ham} of the main text can be transformed by a Jordan-Wigner transformation to an XXZ Hamiltonian. 

Introducing smooth variations of $\mu_n$ and $a_n$ can be used to generate wavepackets of chiral charge to numerically simulate scattering in the lattice fermion model~\cite{vlijm2015quasi,ganahl2012observation}. To understand this, we first apply a gauge transformation (i.e. time-dependent unitary transformation) to set $\mu_n=0$ and transform $\tilde{a}_n(t)=a_n-\int_{-\infty}^t d\tau (\mu_{n+1}-\mu_n)$. In the continuum limit where $\tilde{a}_n$ is slowly varying, the field is Lorentz invariant so that a variation $\tilde{a}_n=f(v t-n)$ is Lorentz invariant. In fact, such a vector potential represents a wavepacket of chiral charge 
that propagates with velocity $v$. We can apply a gauge transformation so that $\partial_t a_n|_{t=0}=0$ and we get $\mu_n=\sum_{m<n}f'(t-m)$ and $a_n=f(t-n)$.
 In our simulations, we use a Gaussian for $f(x)$. To generate Figs.~\ref{fig:trajectory} and \ref{fig:ac_conductance} of the main text, we run simulations on a 640 site system with a 3 site crossover region. We use a Trotter step size of $\tau=2$ which we find is sufficient due to the slow dynamics.

 To measure the ac conductance, we choose a point just past the transition region and measure the time-dependent charge. Specifically, we measure the charge within a Gaussian of width 3 to smooth UV density oscillations. Comparing this profile with the initially inserted packet gives the ac conductance for short frequencies.

\section{Boundary sine-Gordon details}
  Here we elaborate on the mapping from the non-chiral Luttinger liquid to the boundary sine-Gordon model. As outlined in the main text, we begin by folding the system onto the positive half-line and rescaling by the Luttinger parameter, giving $\phi_\L(x)=K_\L^{-1/2}\phi(-x)$ and $\phi_\R(x)=K_\R^{-1/2}\phi(x)$ for $x\geq 0$. 
This transformation sets the Luttinger parameter associated with $\phi_\LR$ to be unity. 
The fields $\phi_\LR(x\sim 0)$, however, change abruptly near $x\sim 0$, so that the boundary condition is written as $K_\R^{-1/2}\partial_x \phi_\R=-K_\L^{-1/2}\partial_x \phi_\L$ and $K_\R^{1/2}\phi_\R=K_\L^{1/2}\phi_\L$. 
This boundary condition can be completely decoupled  by transforming to fields $(\xi,\;\bar\xi)^T=R(\theta)(\phi_\L,\;\phi_\R)^T$ where $R(\theta)$ is an $O(2)$ rotation by an angle $\theta=\sin^{-1}(\sqrt{K_\L}/\Gamma)$, 
with $\xi$ and $\bar\xi$ fields defined as 
\begin{equation}
\left(\begin{matrix} \xi \\ \bar\xi \end{matrix}\right) = %
\Gamma \left(\begin{matrix} K_\L^{-1/2} & K_\R^{-1/2} \\ K_\R^{-1/2} &-K_\L^{-1/2}\end{matrix}\right) %
\left(\begin{matrix} \phi_\L  \\ \phi_\R \end{matrix}\right).
\end{equation}
and likewise with their conjugate momenta $\eta$ and $\bar\eta$ (rescaling by $K^{1/2}_\LR$ instead of $K^{-1/2}_\LR$).
As a reminder, $\Gamma \equiv (K_\L^{-1}+K_\R^{-1})^{-1/2}$.
Under this transformation, the boundary conditions become $\bar\xi(0)=0$ and $\partial_x \xi(0)=0$.
\newcommand{\CFg}{\tilde g}
The effect of a weak impurity $\CFg$ is expected to be limited to low energy modes, 
so that the variation of $K(x\sim 0)$ can be ignored. This leads to the BSG model in Eq.~6 of the main text.

As explained in Eqs.~7 and 8 in the main text, the BSG model is refermionized using a chiral boson on the entire real line. However, in order to equate the charge of the fermionic field with the original BSG field $\xi$, one refolds the field on to the half line:
\begin{equation}
\label{eq:dirac-fermion}
     \psi^\dagger(x)=e^{i \kF x}\chiral{\psi}^\dagger(x)+e^{-i \kF x}\chiral{\psi}(-x)
\end{equation}
such that the total current is
\begin{equation}\label{eq:jF}
    j(x) = (i/2) [\psi^\dagger(x)\partial_x\psi(x)-\mathrm{h.c.}].
\end{equation}

\subsection{Refermionization details}
Though refermionization is standard, its implementation can be prone to subtleties.
Therefore we formalize the refermionization procedure in accordance with Ref.~\onlinecite{von1998bosonization}.
We consider the fields $\xi(x)$ and $\eta(x)$ on the finite wire $x \in (0,L)$. In this case, we can expand each as a Fourier series
\begin{align*}
\xi(x) &= A_0 + \sqrt{2} \sum_{n=1}A_n\, e^{\pi n a/L} \; \cos \frac { \pi n x} L \\
\eta(x) &= B_0 + \sqrt{2}\sum_{n=1}B_n\, e^{\pi n a/L} \; \cos \frac { \pi n x} L 
\end{align*}
where only cosine terms are included since $x>0$ and where $a$ is a soft UV cutoff for the coefficients $A_n$ and $B_n$.
From the canonical commutation relation $[\xi(x),\eta(x')]=i\pi \delta(x-x')$, we find
\begin{align*}
    [A_m, B_n] &= \frac {\pi i}{L} \; e^{-2\pi n a/L} \; \delta_{mn} \\
(a\rightarrow 0) \qquad &= \frac {\pi i}{L} \; \delta_{mn}.
\end{align*}
The chiral field $\chiral\xi(x)=(1/2)[\xi(|x|)+\int_0^x\mathrm dx\, \eta(|x|)]$ becomes 
\begin{equation*}
\chiral\xi(x) = \frac{A_0}{2} + \frac{B_0 x}{2} + \frac 1 {\sqrt{2}} \sum_{n=1} e^{-\pi n a/L}\; \left[A_n \cos \frac { \pi n x} L + \frac{B_n L}{\pi n} \sin \frac { \pi n x} L\right]
\end{equation*}
which obeys 
\begin{align*}
[\chiral\xi(x),\chiral\xi(x')] &= -\frac i 2 \sum_{n=-\infty}^\infty \tan^{-1} \left( \frac{x-x'-2nL} a\right) \\
(a\rightarrow 0) \qquad &= -\frac{i\pi}{4} \, \mathrm{sgn}(x-x').
\end{align*}
This relation lacks the $O(1/L)$ term found in Ref.~\onlinecite{von1998bosonization} due to the inclusion of the zero modes $A_0$ and $B_0$. These zero modes become the Klein factor and $\exp(2\pi N x/L)$ factors in the bosonization identity of Ref.~\onlinecite{von1998bosonization}.

Then replacing $\xi$ with the chiral field in Hamiltonian (Eq.~\ref*{eq:ham2} of the main text) becomes
\begin{equation}
H_\xi = \int_0^L \frac{\mathrm dx}{\pi} \, (\partial_x \chiral\xi(x))^2 + \CFg \cos 2\chiral\xi(0).
\end{equation}
In the noninteracting limit ($\CHg=0$), the equation of motion becomes $\partial_t \chiral\xi=\partial_x \chiral\xi$, demonstrating that $\chiral\xi(x)$ is in fact chiral.

One subtlety in this refermionization procedure is that $\xi(x)$ and $\eta(x)$ are not well-defined operators on the Hilbert space of states. 
This concern was mentioned in Ref.~\onlinecite{von1998bosonization} and remedied with explicit Klein factors. 
Consider for example the unitary transformation $T=\mathrm{exp}(2\pi i N)$ where $N = \pi^{-1} \int_0^L \mathrm dx\, \eta(x)=B_0 L/\pi$ is the chiral fermion number. 
Since $[N,\xi(x)]=i$, the transformed field is $T \xi(x) T^\dagger=\xi(x)+2\pi$ which leads to the equivalence $A_0 \equiv A_0 + 2\pi$. 
In essence, the operator $A_0$ is periodic and only defined on a compact subspace. Since our refermionization prescription (Eq.~\ref*{eq:refermionization} of the main text) only depends on $A_0$ only up to this transformation, the fermion field $\chiral\psi$ is well-defined. In other words, we do not utilize expressions of the form $\mathrm{exp}(ic \chiral\xi(x))$ where $c\notin \mathbb Z$.

\subsection{Andreev Reflection}
Rotating the chiral fermion fields into a Majorana basis given by $\gamma_1=\chiral\psi+\chiral\psi^\dagger$ and $\gamma_2=i(\chiral\psi-\chiral\psi^\dagger)$ leads to free transport for $\gamma_2$ and a decoupled Hamiltonian
\begin{equation}
\label{eq:ham3}
H_{\gamma_1} = \int_{-\infty}^\infty \frac {\mathrm dx}{2\pi} \, \gamma_1(x) i\partial_x \gamma_1(x) + \sqrt{2\EB} \alpha \gamma_1(0)
\end{equation}
for $\gamma_1$.    
The transmission due to the interaction term in Eq.~\ref*{eq:cosF} on the main text can be solved by solving the time evolution of $\gamma_1(x)$ and $\alpha$ using the Heisenberg equation of motion in the frequency domain:
\begin{align*}
\omega \gamma_1(x) &= -[H_{\gamma_1}, \gamma_1(x)] \\
&= i \partial_x \gamma_1(x) - 2\sqrt{2\EB} \alpha \delta(x), \\
\omega \alpha &= -[H_{\gamma_1}, \alpha] \\
&= 2\sqrt{2\EB} \gamma_1(x).
\end{align*}
Integrating the region around $x=0$ gives 
\begin{align*}
\gamma_1(0_+) - \gamma_1(0_-) &= i \sqrt{2\EB} \alpha \\ 
\gamma_1(0_+) + \gamma_1(0_-) &= \frac{2\omega }{\sqrt{2\EB}} \alpha.
\end{align*}
The transmission coefficient is
\begin{align*}
T &= \gamma_1(0_+) / \gamma_1(0_-) \\ 
&= \frac{\omega + i\EB}{\omega-i\EB}
\end{align*}
which is a pure phase.
Transforming back to the fermion basis, and replacing frequency $\omega$ with energy $E$, we calculate the transmission matrix of $\chiral\psi$ to be
\begin{equation*}
t_0 = \frac 1 {iE + \EB} \left( \begin{matrix} -\EB & iE \\ iE & -\EB
\end{matrix}\right).
\end{equation*}
Refolding this system using Eq.~\ref{eq:dirac-fermion}, this transmission matrix leads to a reflection matrix at $x=0$ of 
\begin{align*}
r &=\sigma_x t_0 \\
&= \frac 1 {iE + \EB} \left( \begin{matrix} iE & -\EB \\ -\EB & iE
\end{matrix}\right)
\end{align*}
which results in the amplitudes in Eq.~\ref*{eq:rAndreev} of the main text.

\section{Review of Noise}
Using Eq.~\ref{eq:rAndreev} we can reproduce known formulae for the conductance and shot noise \cite{chamon1997distinct,sandler1998andreev,sandler1999noise}, the latter of which reveals Andreev reflection in the form of the charge of the scattered observables~\cite{muzykantskii1994quantum,de1994doubled}. 
 The noise at high voltages $V \gg \EB$ goes as $P\propto \EB\ll I\sim V/2$ resulting from incoherent normal reflection caused by the impurity that would vanish 
 in the absence of backscattering. Furthermore, the noise power nearly vanishes relative to the current as a result of most of the current being carried 
 by perfect Andreev reflection. In this case, the noise power approaches $P\sim \pi e^2\EB/2 h \sim \delta I$
 where $\delta I$ is the small backscattered component of the current at high voltages. The effective charge of the back-scattered current is $P/2\delta I\sim e/2$.
 On the other hand at low voltages the power goes as $P\sim V^3/12 \EB^2\sim 2 I$, which is consistent with tunneling of electrons 
 as expected from the fraction at $\nu=1$. This is because the effective charge is $P/2\delta I \sim e$, which is twice the high voltage value. This is similar to tunneling in superconductors where the high voltage noise from quasiparticles is twice the result from Cooper pair tunneling at low voltages. 
 
\begin{figure}[h]
\centering
\includegraphics[width=0.6\columnwidth]{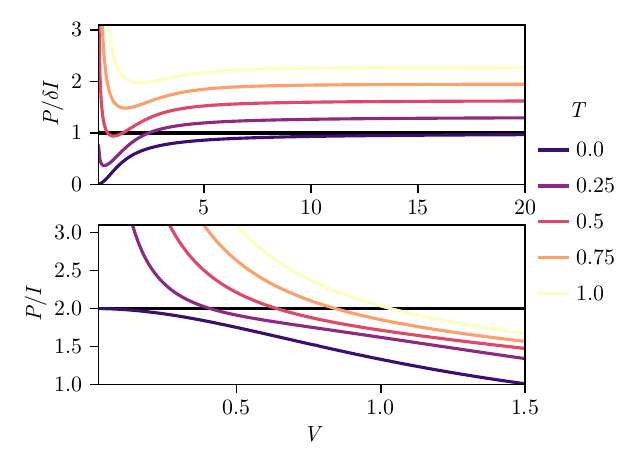}
\caption{\label{fig:noise} Asymptotic behavior of noise $P$ representing $V\rightarrow \infty$ (top) and $V\rightarrow 0$ (bottom) asymptotic behavior. The $T=0$ case matches $P\rightarrow \delta I$ in the large voltage limit and $P\rightarrow 2 I$ in the small voltage limit. The impurity energy is $\EB=1$.}
\end{figure}

The asymptotic behavior of the shot noise is plotted in Fig.~\ref{fig:noise}.

\section{Leads at different temperatures}
An interesting extension of our analysis occurs when the leads are at different temperatures $T_\pm$, which results
in a non-Fermi distribution for the $\psi(x)$ fermion given by the modified Fermi distribution $\tildenF (k)$.
This distribution is derived from noting that $\chiral\psi$ is an exponential of $\xi$ and 
therefore also of $\phi_\pm$ and observing that these bosonic fields have Gaussian distribution. 

The effective fermion field $\chiral\psi$ is expressed in terms of the chiral boson field $\chiral\xi(x)$, which relates to the original chiral fields as
\begin{align*}
    \chiral\xi(x) &= \frac \Gamma {\sqrt{K_-}} \phi_{-,\mathrm c}(x) + \frac \Gamma {\sqrt{K_+}} \phi_{+,\mathrm c}(x)
\end{align*}
where
\begin{align*}
    \phi_{\pm,\mathrm c}(x) &= \frac 1 2 \left[K_\pm^{-1/2} \phi_\pm(x) + K_\pm^{1/2} \int_0^x \mathrm dx'\, \Pi_\pm(x') \right].
\end{align*}
Using the canonical commutation relation $[\phi(x),\Pi(x')]=i \pi \delta(x-x')$ and the folded field definitions $\phi_\pm(x)=\phi(\pm x)$ and $\Pi_\pm(x)=\Pi(\pm x)$, we can show that the two chiral fields 
$\phi_{+,\mathrm c}$ and $\phi_{-,\mathrm c}$ commute with each other (though not with themselves at differing $x$). 
Therefore, products of effective fermions separate as
\begin{align*}
\left\langle \chiral\psi^\dagger(x_1)\, \chiral\psi(x_2) \right\rangle&= \left\langle e^{2i\chiral\xi(x_1)} \, e^{-2i\chiral\xi(x_2)} \right\rangle \\
&= \left\langle e^{(2\Gamma/\sqrt{K_+}) i\phi_{+,\mathrm c}(x_1)} e^{-(2\Gamma/\sqrt{K_+})i\phi_{+,\mathrm c}(x_2)}\right\rangle \left\langle e^{(2\Gamma/\sqrt{K_-})i\phi_{-,\mathrm c}(x_1)} e^{-(2\Gamma/\sqrt{K_-})i\phi_{-,\mathrm c}(x_2)} \right\rangle,
\end{align*}
assuming that far from the impurity $\phi_{-,\mathrm c}$ and $\phi_{+,\mathrm c}$ are independent.
Using the property of bosonic operators that $\langle \mathrm{exp} [\lambda \hat B]\rangle=\mathrm{exp} [\lambda^2 \langle {\hat B}^2 \rangle/2]$, we calculate
\begin{align}
    \left\langle \chiral\psi^\dagger(x_1)\, \chiral\psi(x_2) \right\rangle &= e^{-4\Gamma^2/K_+[\phi_{+,\mathrm c}(x),\phi_{+,\mathrm c}(x')]/2} e^{-4\Gamma^2/K_+\left\langle(\phi_{+,\mathrm c}(x_1)-\phi_{+,\mathrm c}(x_2))^2\right\rangle} \times \left(+\Longleftrightarrow -\right) \nonumber \\
&= \left\langle e^{2 i\phi_{+,\mathrm c}(x_1)}\; e^{-2i\phi_{+,\mathrm c}(x_2)}\right\rangle^{\Gamma^2/K_+} \left\langle e^{2i\phi_{-,\mathrm c}(x_1)} e^{-2i\phi_{-,\mathrm c}(x_2)} \right\rangle^{\Gamma^2/K_-} \nonumber \\
&= \left\langle \psi^\dagger_+ (x_1) \psi_+(x_2)\right\rangle^{\Gamma^2/K_+} \left\langle \psi^\dagger_- (x_1) \psi_-(x_2) \right\rangle^{\Gamma^2/K_-}.\label{eq:fermion-correlator}
\end{align}
where $\psi_\pm(x) = \mathrm{exp}[ 2i\phi_\pm(x) ]$. If $\psi_\pm$ is described by a Fermi-Dirac distribution $\nF$ with temperature $T_\pm$, these correlators are given by the Fourier transform of $\nF(k)$:
\begin{equation*}
\left\langle \psi^\dagger_\pm (x+r) \psi_\pm(x)\right\rangle = -\frac{i\pi T_\pm}{\sinh \pi T_\pm r}
\end{equation*}
assuming translation invariance far from the boundary. Using this expression and Eq.~\ref{eq:fermion-correlator}, we find an effective distribution

\begin{align}
    \tildenF(k) &=\int \mathrm dx\, e^{i k x}\left(\frac{-i\pi T_+}{\sinh{\pi T_+ x}}\right)^{\Gamma^2/K_+}\left(\frac{-i\pi T_-}{\sinh{\pi T_- x}}\right)^{\Gamma^2/K_-}, \nonumber \\
    &=\int_0^\infty \mathrm dx \sin{(k x)} \, \Phi(x),
\end{align}
where 
\begin{equation*} 
 \Phi(x)=\left(\frac{T_\R}{\sinh{\pi T_\R x}}\right)^{\Gamma^2/K_\R}\left(\frac{T_\L}{\sinh{\pi T_\L x}}\right)^{\Gamma^2/K_\L}.
\end{equation*} 
This formula reduces to the Fermi distribution for $T_\L=T_\R$. 

\begin{figure}
\centering
\includegraphics[width=0.6\columnwidth]{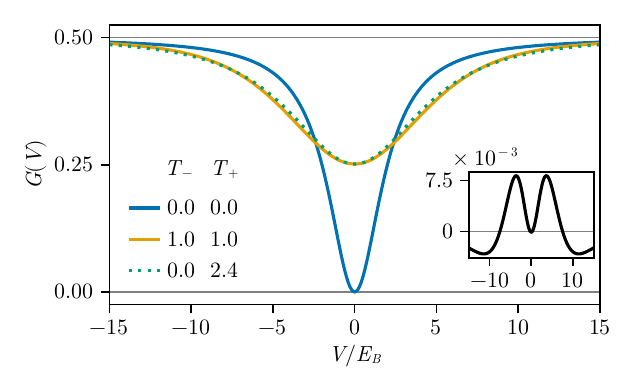}
\caption{\label{fig:G_V} Voltage-dependent conductance for various temperatures demonstrating approximate equivalence between asymmetric and symmetric lead temperatures. Inset shows absolute difference between $T_\L=0$, $T_\R=2.4$ conductance and $T_\L=1.0$, $T_\R=1.0$.}
\end{figure}

The current, conductance and shot noise can now be computed by simply substituting $\tildenF$ for $\nF$. For the conductance, we can use the Eq.~\ref{eq:rAndreev} from the main text to get 
\begin{align*}
    G(V) &=\frac{e^2}{2 h} \left[ 1-\frac \EB 2\int_0^\infty \mathrm dx \,x\,\Phi(x)\,e^{-\EB x}\cos{(V/2) x}\right].
\end{align*}
We plot the results of this function, numerically integrated with a Gauss-Kronrod quadrature formula~\cite{quadgk}, in Fig.~\ref{fig:G_V}. The result shows a small difference in the conductance when the leads are at different temperatures compared to the conductance with an effective single temperature.

\bibliography{bib.bib}